\documentclass[12pt]{article} 

\usepackage{amsmath,latexsym}

\newcommand{\RR}{\hbox{I\kern-.2em\hbox{R}}}
\begin{document}

	\begin{center}
		{\LARGE \bf Thermodynamic stability of blackbody radiations at low temperatures 
		}
		\\ 
		\vspace{1cm} 
		{\large E. S. Moreira Jr.}
		\footnote{E-mail: moreira@unifei.edu.br}   
		\\ 
		\vspace{0.3cm} 
		{\em Instituto de Matem\'{a}tica e Computa\c{c}\~{a}o,}  
		{\em Universidade Federal de Itajub\'{a},}   \\
		{\em Itajub\'a, Minas Gerais 37500-903, Brazil}
		
		\vspace{0.3cm}
		{\large November, 2026}
	\end{center}
	\vspace{0.5cm}


	\abstract{ 
		In fact a more accurate title for this paper would be
		``Thermodynamic stability of blackbody radiations at low temperatures and/or small cavities'' since we will be dealing with the regime $kTV^{1/3}/\hbar c\ll 1$, where $T$ is the temperature of 
		a massless quantum field of arbitrary spin encapsulated in a
		cavity of volume $V$. By studying heat capacities and compressibilities, our main result is that stable thermodynamic equilibrium requires strictly positive vacuum (``Casimir'') energies. That is to say, negative or vanishing vacuum energies necessarily lead to thermodynamic instability.
		Other thermodynamics quantities are determined and their relationships are checked. It is argued that our results can be extended to models of compact universes. 
		}

	\section{Introduction}
	\label{introduction}
	It is hard to begin a paper about blackbody radiation without recalling that it is the protagonist in the birth of quantum mechanics. It is essential for live, and it is generated inside our laboratories and accelerators at all temperatures. On a cosmic scale, it permeates the entire universe at a temperature of approximately 3 K \cite{pen65}. All these facts are well known.
	
	In a recent study \cite{mor24}, among other things, we have investigated thermodynamic stability of the electromagnetic blackbody radiation at the regime 
	where $kTV^{1/3}/\hbar c\gg 1$, which is called the high temperatures $T$ (and/or large volumes $V$) regime. Two perfect conducting cavities were examined: a spherical shell and a cubic box. The criteria for stable thermodynamic equilibrium, namely that heat capacities and compressibilities must be  positive quantities \cite{cal85}, have shown to be fulfilled by 
	electromagnetic blackbody radiation 
	in the spherical shell but not 
	in the cubic box \cite{mor24}.	We then decided to extend the investigation to address the opposite regime
	\begin{equation}
	\label{ltregime}
	\frac{kTV^{1/3}}{\hbar c}\ll 1
\end{equation}
		and in so doing we realized that the framework was the same for massless quantum fields of all spins. It turns out that the present paper concerns the thermodynamics of all types of blackbody radiation satisfying equation (\ref{ltregime}). Our main goal is to study thermodynamic stability. In the following two sections we address thermodynamic in a cavity, presenting our results. In the last section we consider some applications
		and extend the results to $\RR\times S^{d}$ universes.

	\section{Thermodynamic stability}
	\label{ts}	
	Thermodynamic criteria for stability arise from the requirement that the entropy $S$ is a local maximum \cite{cal85}. It results that the heat capacity at constant volume $C_{V}$ and the isothermal compressibility $\kappa_{T}$ must be positive, i.e.,
	\begin{equation}
		C_{V}:=\left(\frac{\partial U}{\partial T}\right)_{V}>0, \hspace{2.0cm} 
		\kappa_{T}:=-\frac{1}{V}\left(\frac{\partial V}{\partial P}\right)_{T}>0
		\label{ck1}
	\end{equation}
    for a thermodynamic system of pressure $P$, volume $V$, temperature $T$, and internal energy $U$. The first and second inequalities stand for thermal and mechanical stability, respectively.
    Considering further the definitions of the coefficient of thermal expansion $\alpha$, the heat capacity at constant pressure
    \footnote{Recall that $H:=U+PV$.}
    $C_{P}$,
    and the adiabatic compressibility $\kappa_{S}$,
    \begin{equation}
    	\alpha:=\frac{1}{V}\left(\frac{\partial V}{\partial T}\right)_{P}, \hspace{2.0cm} 
    	C_{P}:=\left(\frac{\partial H}{\partial T}\right)_{P}, \hspace{2.0cm}
    	\kappa_{S}:=-\frac{1}{V}\left(\frac{\partial V}{\partial P}\right)_{S}
    	\label{akapa}
    \end{equation}
	along with the identities
	\begin{equation}
		C_{P}-C_{V}=\frac{TV\alpha^{2}}{\kappa_{T}},
		\hspace{2.0cm}
		\frac{C_{V}}{C_{P}}=\frac{\kappa_{S}}{\kappa_{T}},
		\label{identity1}
	\end{equation}
	it follows then that both heat capacities and both compressibilities must be positive in a stable thermodynamic system \cite{cal85}.
	
	\section{Blackbody radiations at low temperatures}
	\label{bb}
	According to statistical mechanics lore, one begins with  Helmholtz's free energy \cite{pat72,hua87}
	\begin{equation}
		F=U_{0}\pm\frac{1}{\beta}\sum_{\bf k}
		\ln\left(1\mp e^{-\beta\hbar\omega_{{\bf k}}}\right)
		\label{F}
	\end{equation}
	where $U_{0}$ is the vacuum energy and $\beta:=1/kT$ is the inverse temperature. The upper and lower signs are for bosons and fermions quanta, respectively. We notice that 
	$\omega_{{\bf k}}^{2}/c^2\propto V^{-2/3}$ 
	are eigenvalues of minus the Laplacian operator, $-\Delta$, which are determined by the cavity's geometry and boundary conditions on the walls. Now, taking into account equation (\ref{ltregime}), the smallest eigenvalue $\omega_{{\bf 0}}^{2}/c^2$ with degeneracy $g$, and defining $\Theta:=\hbar \omega_{{\bf 0}}/k$, it turns out that equation (\ref{F}) becomes
	\footnote{In fact for bosons, if $\omega_{{\bf 0}}=0$, a ``zero-mode'' divergence appears in equation (\ref{F}) which requires regularization. Nevertheless this detail will not change our conclusions.}
	\begin{equation}
		F=U_{0}-gkT  e^{-\Theta/T}
		\label{f}
	\end{equation}
	where, of course, exponential smaller terms have been omitted and it will be so along the whole text. 
	
	On dimensional grounds we also have that
	\begin{equation}
		U_{0}=\lambda \frac{\hbar c}{V^{1/3}}.
		\label{casimir}
	\end{equation}
	Since the seminal work of Casimir \cite{cas48},  values of the dimensionless constant $\lambda$ in equation (\ref{casimir}) have been found for many fields, geometries, and boundary conditions
	by using a variety of regularization and renormalization procedures \footnote{For reviews on the Casimir effect, see e.g. references \cite{mil01,bor01}.}. It is worth remarking that $\lambda$ can be negative, null, or positive. Then, when $\lambda=0$, the exponential small term in equation (\ref{f}) becomes the leading contribution.
	
	The entropy $S=-(\partial_{T} F)_{V}$ follows from equation (\ref{f})
	\begin{equation}
		S=gk\frac{\Theta}{T}\left(1+\frac{T}{\Theta}\right)e^{-\Theta/T}
	\label{s}	
	\end{equation}
which, as expected, fulfills the third law of thermodynamics, i.e., $S\rightarrow 0$ as $T\rightarrow 0$. One can now read $C_{V}=T(\partial_{T} S)_{V}$ from $S$ in equation (\ref{s})
	\begin{equation}
	C_{V}=gk\left(\frac{\Theta}{T}\right)^{2} e^{-\Theta/T}
	\label{cv}	
\end{equation}	
resulting that $C_{V}$ is clearly positive as thermodynamic stability demands. It is a simple matter to obtain the internal energy $U=F+TS$
from equations (\ref{f}) and (\ref{s}), namely
\begin{equation}
	U=U_{0}+gk\Theta  e^{-\Theta/T}.
	\label{u}
\end{equation}
Note that as $T\rightarrow 0$ only the Casimir energy in equation (\ref{casimir})	is left behind in equation (\ref{u}), and that by using the first definition in equation (\ref{ck1}), equation (\ref{cv}) is reproduced consistently. Now, by noticing equation (\ref{f})  
and that $\Theta\propto V^{-1/3}$,
$P=-(\partial_{V} F)_{T}$ yields the following equation of state
\begin{equation}
	PV=\frac{U}{3}.
	\label{p}
\end{equation}
Perhaps it may be pedagogical to obtain equation (\ref{p}) in another way, which is to consider $S$ as a function of $U$ and $V$, and to use
\begin{equation}
P=T\left(\frac{\partial S}{\partial V}\right)_{U}.	
\label{p2}
\end{equation}
Indeed, noting equations (\ref{s}) and (\ref{u}), it follows that
$$S=-\frac{U-U_{0}}{\Theta}\left[\ln \left(\frac{U-U_{0}}{gk\Theta}\right)-1 \right]$$
which through equation (\ref{p2}) leads to equation (\ref{p}).

Before turning to the calculation of other quantities
it is worth mentioning that, when $U_{0}\neq 0$, the sign of $P$ in equation (\ref{p}) is determined by the sign of $U_{0}$ in equation (\ref{casimir})
(i.e., by the sign of $\lambda$). Then, a positive Casimir energy  inflates the cavity whereas a negative Casimir energy shrinks it
[cf. equation (\ref{p2})]. Now, if $\lambda=0$,  corresponding to a vanishing $U_{0}$, equations (\ref{u}) and (\ref{p}) tell us that $P$ is exponentially small and positive, therefore the cavity tends to inflate.

To determine $\alpha$ in equation (\ref{akapa}) we calculate $(\partial_{T}V)_{P}$ by considering $P$ constant in equation (\ref{p}) and taking $\partial_{T}$ of both sides noting equation (\ref{u}). After some algebra we get
\begin{equation}
\frac{4}{3}\left(\frac{\partial V}{\partial T}\right)_{P}
\left[U_{0}-\frac{gk\Theta^2}{4T}\left(1-\frac{4T}{\Theta}\right)e^{-\Theta/T}\right]
=gkV\left(\frac{\Theta}{T}\right)^2 e^{-\Theta/T},
\nonumber
\end{equation}
leading to 
\begin{eqnarray}
&&\alpha=\frac{3gk}{4U_{0}}\left(\frac{\Theta}{T}\right)^2 e^{-\Theta/T}, \hspace{1.0cm}	U_{0}\neq 0,
\label{a1}
\\
&&
\alpha=-\frac{3}{T}\left(1-\frac{4T}{\Theta}\right)^{-1},
\hspace{1.0cm}	U_{0}=0.
\label{a2}
\end{eqnarray}
Looking at equation (\ref{a1}) one sees that $\alpha<0$ if $U_{0}<0$,
resulting that by keeping $P$ constant the cavity shrinks as it warms
[cf. equation (\ref{akapa})]. And this is also the case for a vanishing Casimir energy according to equation (\ref{a2}). In contrast, when $U_{0}>0$ and $P$ is kept constant, the cavity inflates as it warms, which is the behavior expected for an ordinary gas.

In order to calculate $\kappa_{T}$ in equation (\ref{ck1}), we begin by considering $(\partial_{V}P)_{T}$ from which we can read  $(\partial_{P}V)_{T}$, i.e., $1/(\partial_{V}P)_{T}$. To calculate 
$(\partial_{V}P)_{T}$ we use again equations (\ref{p}) and (\ref{u}). 
Thus
\begin{equation}
	\left(\frac{\partial P}{\partial V}\right)_{T}=-\frac{4}{9V^2}
	\left[U_{0}-\frac{gk\Theta^2}{4T}\left(1-\frac{4T}{\Theta}\right)
	e^{-\Theta/T}\right],
	\nonumber
\end{equation}
resulting in
\begin{eqnarray}
	&&\kappa_{T}=\frac{9V}{4U_{0}}\left[1+\frac{gk\Theta^2}{4TU_{0}}
	\left(1-\frac{4T}{\Theta}\right) e^{-\Theta/T}\right], \hspace{1.0cm}	U_{0}\neq 0,
	\label{k1}
	\\
	&&
	\kappa_{T}=-\frac{9TV}{gk\Theta^2}\left(1-\frac{4T}{\Theta}\right)^{-1}e^{\Theta/T},
	\hspace{2.5cm}	U_{0}=0.
	\label{k2}
\end{eqnarray}
By taking into account equation (\ref{ck1}), an exam of  equations
(\ref{k1}) and (\ref{k2}) clearly leads to:
\begin{eqnarray}
\texttt{Thermodynamic stability}&\implies& U_{0}>0.
\label{ts}	
\end{eqnarray}

For the sake of completeness, one may obtain $C_{P}$ and $\kappa_{S}$ from the identities in equation (\ref{identity1}) or, of course, by using their very definitions in equation (\ref{akapa}). Either way, it results that
\begin{eqnarray}
	&&C_{P}=C_{V}[1+\cdots], \hspace{3.7cm}	U_{0}\neq 0,
	\label{cp1}
	\\
	&&
	C_{P}=-4gk\frac{\Theta}{T}\left(1-\frac{4T}{\Theta}\right)^{-1}e^{-\Theta/T},
	\hspace{1.0cm}	U_{0}=0,
	\label{cp2}
\end{eqnarray}
where $C_{V}$ is given in equation (\ref{cv}) with ellipsis denoting a positive exponentially small correction, and
\begin{eqnarray}
	&&\kappa_{S}=\frac{9V}{4U_{0}}\left[1-\frac{gk\Theta}{U_{0}}
	e^{-\Theta/T}\right], \hspace{1.0cm}	U_{0}\neq 0,
	\label{ks1}
	\\
	&&
	\kappa_{S}=\frac{9V}{4gk\Theta}e^{\Theta/T},
	\hspace{3.0cm}	U_{0}=0.
	\label{ks2}
\end{eqnarray}
Before turning to our conclusions, it should be noticed that for a non vanishing Casimir energy, $C_{V}\approx C_{P}$ and $\kappa_{T}\approx\kappa_{S}$.

	\section{Conclusions}
	\label{summary}
The initial purpose of this paper was to extend reference \cite{mor24} to low temperatures and/or small cavities. It turned out however that the outcome has become wider addressing thermodynamic stability of any blackbody radiation in compact spaces. 
By going through well known criteria  for thermodynamic stability, the paper has shown that positiveness of the system's Casimir energy is a necessary condition for the blackbody thermodynamic stability [cf. equation (\ref{ts})].	
A few consequences of this fact will be addressed below. 

\subsection{Cubic boxes and spherical shells} 

In other to compare thermodynamic stability across regimes of temperature (and/or cavity size), let us briefly see the study of mechanical stability for massless scalar radiation in a cubic box of volume V at the regime  
[cf. equation (\ref{ltregime})]
	\begin{equation}
	\label{rtregime}
	\frac{kTV^{1/3}}{\hbar c}\gg 1.
\end{equation}
The Helmholtz free energy can be obtained, e.g., from the general formula in reference \cite{dow84},
\begin{equation}
	F=-\frac{\pi^{2}}{90}\frac{(kT)^{4}}{(\hbar c)^{3}} V
	\pm\frac{3\zeta(3)}{4\pi}\frac{(kT)^{3}}{(\hbar c)^2}V^{2/3},
	\label{fcb}
\end{equation}
where smaller  corrections have been omitted and upper and lower signs are for Dirichlet or Neumann boundary conditions on the cubes's sides, respectively. It quickly follows that
\footnote{Recall that $P=-(\partial_{V} F)_{T}$, and see equation (\ref{ck1}).}
\begin{eqnarray}
	\hspace{-1.0cm} P=\frac{\pi^{2}}{90}\frac{(kT)^{4}}{(\hbar c)^{3}}
	\mp\frac{\zeta(3)}{2\pi V^{1/3}}\frac{(kT)^{3}}{(\hbar c)^2}, 
	\hspace{1.0cm} 
	&&
	\kappa_{T}=\mp\frac{6\pi}{\zeta(3)}\frac{(\hbar c)^{2}}{(kT)^{3}}V^{1/3}.
	\label{sp2}
\end{eqnarray}
Thus, noticing equations (\ref{ck1}) end (\ref{p}), one has that the thermodynamic stability criteria hold for Neumann's cube, but not for  Dirichlet's cube, at high temperatures. Now, it is well known that the Casimir energy in Dirichlet and Neumann cubes are both negative
\footnote{More precisely, $\lambda$ in equation (\ref{casimir}) is approximately -0.016 and -0.29 for the Dirichlet and Neumann cubic boxes, respectively.} \cite{amb83}. Then, considering equation (\ref{ts}), the whole scenario is as follows. At low temperatures (and/or small cavities), scalar blackbody radiation is not thermodynamically stable either in the Dirichlet or in the Neumann cubic boxes. At high temperatures (and/or large cavities),
scalar blackbody radiation remains thermodynamically unstable in  Dirichlet's cubic box, but now satisfies the thermodynamic stability criteria in Neumann's.

As mentioned in section \ref{introduction}, by addressing perfect conducting cavities,
reference \cite{mor24} has shown that electromagnetic blackbody radiation
at the regime in equation (\ref{rtregime})
satisfies the criteria for stable thermodynamic equilibrium [cf. equation (\ref{ck1})] in a spherical shell but not in a cubic box \footnote{In a cubic box $\kappa_{T}<0$ \cite{mor24}.}. As the corresponding Casimir energies in  these cavities are both positive, and indeed comparable \cite{amb83}, the condition for thermodynamic stability at low temperatures (and/or small cavities) in equation (\ref{ts}) is fulfilled by both geometries. 

We could carry on studying other fields (such as spinors) in a variety of 
cavities and corresponding
boundary conditions. But these examples suffice.

\subsection{Einstein universes: $\RR\times S^{d}$}

Perhaps the simplest model of blackbody radiation living in an Einstein universe is a massless scalar field at finite temperature $T$ in $\RR\times S$, i.e., blackbody radiation on a circle \cite{mor20}. The geometry of $\RR\times S^{3}$, where $S^{3}$ is the $3$-sphere of radius $a$ and volume $\propto a^{3}$, is a solution of Einstein's equations and could be the geometry of our universe at some point of its history. For a conformally invariant scalar field in $\RR\times S^{3}$, at the regime of low $T$ (and/or small $a$), equations (\ref{u}) and (\ref{p})
equally apply with $U_{0}=\hbar c/240 a$ and $\Theta\propto 1/a$ \cite{for75,dow77}. At high $T$ (and/or large $a$) equation (\ref{p}) still holds and $P$ is given as in equation (\ref{sp2})
with the correction to the Planck term replaced by a positive exponentially small term that vanishes when $a\rightarrow \infty$.

The conformally invariant scalar field
picks up a particular value of the curvature coupling parameter $\xi$ (with the Ricci curvature), which is determined by the dimension of $S^d$ \cite{dav82}. In $d=3$, such a value is $1/6$.
It has been recently shown that, for massless scalar fields in $\RR\times S^{3}$, only the conformal coupling $\xi=1/6$ is consistent with stable thermodynamic equilibrium at 
all $T$ and for all $a$ \cite{mor26}. 
This fact might suggest that the same happens in arbitrary $d$.
In other words, it might suggest that conformally invariant scalar fields would be thermodynamically stable in every $\RR\times S^{d}$. Nevertheless, according to the results in the present work, this is incorrect.

References \cite{ozc06,her08} have thoroughly studied 
$U_{0}$ for conformally invariant scalar fields in $\RR\times S^{d}$.
They have shown that for even $d$, $U_{0}=0$. And that for $d=1$, $d=5$,
and for other odd
\footnote{In fact, for odd $d$, $U_{0}$ oscillates in sign.}
values $d>7$, $U_{0}<0$. In 
accordance with eq. (\ref{ts}), all these cases correspond to thermodynamic instability. It is safe to say that conformal invariance does not imply thermodynamic stability.

Thus, the case  $\RR\times S^{3}$ seems to be alone among the dimensions. One might even conjecture that the result reported in reference \cite
{mor26}, namely, that  conformally invariant scalar fields are thermodynamic stable for all $T$ and for all $a$ is a special case.
But this, of course, requires further investigation which is beyond the scope of this paper.

\vspace{1cm}
\noindent{\bf Acknowledgements} -- 
Work partially supported by
``Funda\c{c}\~{a}o de Amparo \`{a} Pesquisa do Estado de Minas Gerais'' (FAPEMIG).

\end{document}